\documentclass[journal]{IEEEtran}
\IEEEoverridecommandlockouts
\usepackage{multirow}
\usepackage{booktabs}
\usepackage{graphicx}
\usepackage{cite}
\usepackage{amsmath,amssymb,amsfonts,bbm}

\usepackage{algpseudocode}
\usepackage{graphicx}
\usepackage[caption=false, font=footnotesize]{subfig}
\usepackage{balance}
\usepackage{textcomp}
\usepackage[normalem]{ulem}
\usepackage{xcolor}
\usepackage[linesnumbered,ruled,vlined]{algorithm2e}
\SetKwInput{KwInput}{Input}                
\SetKwInput{KwOutput}{Output}              

\def\BibTeX{{\rm B\kern-.05em{\sc i\kern-.025em b}\kern-.08em
    T\kern-.1667em\lower.7ex\hbox{E}\kern-.125emX}}

\begin{document}
\bstctlcite{IEEEexample:BSTcontrol}

\font\myfont=cmr12 at 21pt
\title{\myfont Towards Ultra-Reliable 6G in-X Subnetworks: Dynamic Link Adaptation by Deep Reinforcement Learning
}

\author{
    Fateme Salehi, \IEEEmembership{Member, IEEE}, 
    Aamir Mahmood, \IEEEmembership{Senior Member, IEEE},
    Mustafa Ozger, \IEEEmembership{Member, IEEE},
    Sarder Fakhrul Abedin, \IEEEmembership{Senior Member, IEEE},
    Kyi Thar, 
    and Mikael Gidlund, \IEEEmembership{Fellow, IEEE}
\vspace{-15pt}
\thanks{F. Salehi {\emph{(Corresponding author)}} and M. Ozger are with the Department of Electronic Systems, Aalborg University, Frederikskaj 12, 2450 København SV, Denmark (e-mail: \{fsalehi, mozger\}@es.aau.dk). A. Mahmood, K. Thar, and M. Gidlund are with the Department of Computer and Electrical Engineering,  Mid Sweden University, Holmgatan 10, 851 70 Sundsvall, Sweden (e-mail: \{aamir.mahmood, kyi.thar, mikael.gidlund\}@miun.se). S. F. Abedin is with the Department of Computer Science \&
Industrial Systems,  RISE Research Institutes of Sweden, Isafjordsgatan 22, 16440 Kista, Sweden (e-mail: sarder.fakhrul.abedin@ri.se).}
}

\maketitle

\begin{abstract}
6G networks are generally composed of subnetworks expected to meet ultra-reliable low-latency communication (URLLC) requirements for mission-critical applications such as industrial control and automation. An often overlooked aspect in URLLC is burst/consecutive packet outages, which can destabilize control loops and compromise safety in in-factory environments. Hence, this work proposes a link adaptation framework to satisfy probabilistic reliability constraints using the soft actor-critic (SAC)-based deep reinforcement learning (DRL) algorithm that jointly optimizes reliability and energy efficiency (EE) under dynamic channel and interference conditions. Unlike prior work focusing on average reliability, our method explicitly targets reducing consecutive outages through adaptive control of transmit power and blocklength based on delayed channel quality and the number of consecutive outages, along with link availability observed up to the previous timestep. The joint chance-constrained optimization problem is formulated under finite blocklength with resource, reliability, and availability constraints, balancing reliability and EE. Simulation results show that the proposed method outperforms the baseline DRL algorithms, reducing outage bursts while consuming only around 37\% of the transmission cost required by a full/maximum resource allocation policy in the evaluated scenario. The framework also supports flexible trade-off tuning between reliability and EE by adjusting reward weights, making it adaptable to diverse industrial requirements.

\end{abstract}


\section{Introduction}
\IEEEPARstart{U}{ltra}-reliable low-latency communication (URLLC) is a key service class in 5G and emerging 6G networks, targeting block error rates as low as $10^{-5}$ and air-interface delays below one millisecond \cite{extremeURLLC}. These guarantees are essential for mission-critical industrial applications such as real-time motion control, collaborative robotics, and closed-loop process automation. To reduce infrastructure costs and simplify deployment, many industrial Internet-of-things (IIoT) systems are increasingly operating in the unlicensed Industrial, Scientific, and Medical (ISM) spectrum. 
Extending URLLC capabilities to unlicensed spectrum through 5G New Radio-Unlicensed (NR-U) can enhance IIoT connectivity and enable cost-effective deployments of industrial wireless networks. However, the ISM band is heavily utilized by diverse technologies such as Wi-Fi, Bluetooth, Zigbee, and proprietary industrial protocols, resulting in significant cross-technology interference (CTI) \cite{simone}. This interference poses substantial challenges for 5G NR-U deployments, which must ensure coexistence with other technologies while maintaining the stringent performance requirements of mission-critical URLLC applications. 
6G in-X subnetworks, where ‘X’ denotes the specific physical domains or environments hosting the subnetwork, such as robots, vehicles, factories, and hospitals, demand extreme communication requirements in terms of throughput, reliability, and latency \cite{6G_vision}. 
This calls for interference-aware NR-U coexistence mechanisms beyond generic interference coping strategies.

\subsection{Conventional Interference Mitigation}
Managing CTI in a wireless network is complex, particularly in shared-spectrum environments such as the ISM band. Various approaches have been developed to address this issue. Interference avoidance strategies \cite{interf_avoid} focus on proactively preventing interference through techniques such as dynamic channel allocation, time-division access, and spatial separation (e.g., directional antennas or beamforming) to minimize overlaps between transmissions. The work in \cite{subnet_interf} presents environment-aware mechanisms for interference management in 6G in-X subnetworks via dynamic channel allocation.
Cognitive radio techniques \cite{cognitive} rely on dynamic spectrum access and intelligent spectrum sensing to detect and use underutilized frequency bands, allowing devices to opportunistically avoid interference and adapt to changing spectrum conditions.
Interference mitigation techniques \cite{interf_mitigat} aim to reduce the impact of interference through approaches such as robust signal processing techniques, interference cancellation, and power control.

This calls for interference-aware NR-U coexistence mechanisms beyond generic interference coping strategies.  
To this end, novel schemes propose leveraging interference prediction methods as a foundation for link adaptation (LA). The works in \cite{KDE, DTMC, DTMC_GC} employ model-based techniques such as kernel density estimation and discrete state space discrete-time Markov chain to analyze tail statistics for predicting future interference power, demonstrating their suitability for URLLC applications.
In \cite{EVT, EVT_transformer}, two probabilistic prediction methods are proposed to model the tail behavior of interference using extreme value theory (EVT) and a transformer-based architecture, respectively.
In the dynamic landscape of wireless communication systems, the high complexity of model-based methods and the significant overhead of collecting environmental parameters necessitate a new methodology that can encapsulate all dynamic aspects of the system (such as the radio channel and interference power) into a real-time state without relying on predefined models.
As a result, there is a pressing need for innovative and efficient resource allocation strategies capable of adapting to these dynamic factors to ensure robust and reliable communication performance. 

\subsection{RL for Resource Management}
Recently, numerous works have leveraged deep reinforcement learning (DRL) to address resource allocation and decision-making challenges in wireless networks. 
The authors in \cite{MA_DRL} proposed a user scheduling and downlink power control mechanism for distributed resource management and interference mitigation in multi-cell wireless networks using multi-agent DRL.
In \cite{RF/VLC}, the authors introduced a heterogeneous radio frequency/visible light communication industrial network architecture and proposed a deep post-decision state-based experience replay and transfer RL algorithm for a joint uplink and downlink resource management problem.
In \cite{DQN_IIoT}, a deep Q-network-based scheme was proposed to address bandwidth utilization and energy efficiency (EE) in a networking graph-based IIoT system.

The authors in \cite{opt_DRL} introduced an optimization theory-based DRL framework for the design of a wireless networked control system (WNCS) to minimize the power consumption.
The work in \cite{wncs_scheduling} considered a joint uplink and downlink scheduling problem of a fully distributed WNCS with a limited number of frequency channels. It formulated the optimal transmission scheduling problem into a Markov decision process (MDP) problem with a countable state space for achieving the minimum expected total discounted cost.
The authors in \cite{MARL} propose a resource management framework for 6G in-X subnetworks based on multi-agent DRL, which can extract inter-subnetwork interference relationships from the received signal strength indicator (RSSI) without requiring instantaneous channel gain between subnetworks.
\textit{However, these studies typically focus on average performance metrics such as throughput, delay, or reliability, and overlook the temporal structure of transmission failures, which can have critical implications for delay-sensitive control applications.}

\subsection{Summary of Contributions}
In industrial wireless control systems, the occurrence of sequential transmission failures, referred to as consecutive outages, can be more detrimental than the high probability of a mean outage. Although a link may meet average reliability constraints, clustered outages can violate real-time deadlines and destabilize control loops. 
Despite the importance of consecutive outages for reliability performance, to the best of our knowledge, this critical aspect has not been investigated in existing research \cite{MA_DRL,RF/VLC,DQN_IIoT,opt_DRL,wncs_scheduling,MARL} for radio resource management (RRM). This metric is particularly significant for industrial control and cyber-physical systems \cite{Onboard}, where resilience to external disturbances is crucial for meeting real-time task deadlines. For instance, in \cite{Santucci}, the probability of $n$ consecutive outages/dropouts is derived for co-designing WNCSs. To address these critical gaps, this paper tackles the challenge posed by the consecutive outage issue in URLLC for in-factory subnetworks impaired by interference. Specifically, we propose an energy-efficient resource allocation algorithm using DRL to mitigate consecutive outages while guaranteeing high reliability and link availability requirements. 
By ensuring high reliability and resource efficiency, this research paves the way for robust communication infrastructures capable of meeting the stringent demands of Industry 4.0 and beyond \cite{industry_4, zeb2024towards}.

The main contributions of this paper are summarized as follows:
\begin{itemize}
\item We consider various link quality metrics, including reliability, link availability, and consecutive outages, and incorporate them into a joint optimization framework to balance a trade-off between reliability and resource consumption.
\item We propose a novel approach based on soft actor-critic (SAC) to dynamically adjust transmission power and blocklength based on the delayed interference and channel gain, the number of observed consecutive outages, and the experienced link availability, optimizing both consecutive outages and EE.
\item The effectiveness of the proposed intelligent DRL-based algorithm is evaluated against benchmarks through comprehensive simulations under dynamic network conditions and various system parameters. 
\end{itemize}

The rest of the paper is organized as follows. Section \ref{sec.sys} presents the system model, link quality metrics, and problem formulation. 
Section \ref{sec.method} presents the DRL-based resource allocation method for the URLLC application. 
Section \ref{sec.res} discusses the experimental results and compares the proposed method with 
other DRL schemes. Section \ref{sec.con} concludes the paper.

\section{System Model and Problem Formulation}\label{sec.sys}
\begin{figure}[!t]
    \centering
    \includegraphics[width=0.85\linewidth]{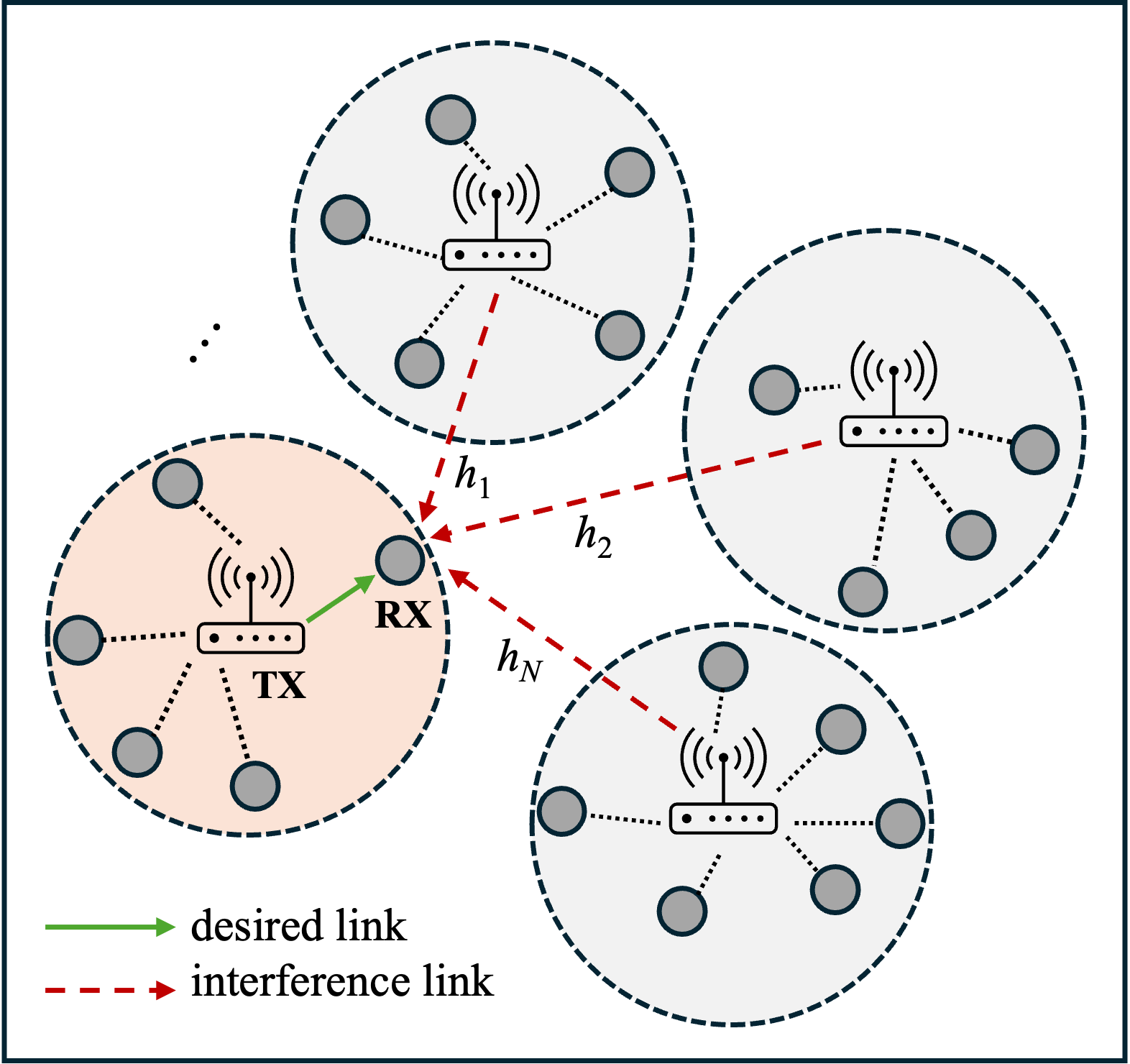}
    \vspace{-2mm}
    \caption{System model: IoT network representing the coexistence of the desired subnetwork with $N$ interfering subnetworks.}
    \label{fig.system}
    \vspace{-5mm}
\end{figure}

The system model shown in Fig. \ref{fig.system} captures the coexistence of multiple in-factory subnetworks in a shared spectrum environment and emphasizes the need for inter-subnetwork interference management and robust communication between the desired transmitter (TX) and receiver (RX). In this model, we consider an IoT network focusing on a URLLC link in a desired subnetwork operating in the presence of $N$ interfering subnetworks. The green solid arrow from TX to RX indicates the desired link in the downlink scenario, while the red dashed arrows from the $k$th interfering subnetwork, $\mathrm{I}_k, k=\{1,\cdots, N\}$, to RX show the interference links. We assume that for intra-subnetwork communication, orthogonal radio resources are allocated to the nodes, ensuring that there is no intra-subnetwork interference. Furthermore, it is assumed that all subnetworks operate independently, with no cooperation among them.

\subsection{System Model}
URLLC transmissions usually occur over mini-slots. Therefore, time is discretized into mini-slots $t$. At a given mini-slot $t$, the $k${th} interferer is either idle or transmitting with power $p_k(t)$. The probability of a new transmission of each interferer is denoted by $\mu$, which is defined as the activation factor. The message duration $l$ is assumed to be the same for all transmissions. A single-antenna Rayleigh block-fading channel model is adopted, given that the channel coherence time in a typical wireless environment is much larger than the mini-slot duration. The aggregate interference at the location of the desired RX at time $t$ is the sum of the receiving powers from all the transmitting nodes,
\begin{equation}
\label{eq.sumInterf}
    I(t)=\sum_{k=1}^N p_k(t) h_k^2(t) \delta_k(t)\ , 
\end{equation}
where $h_k(t)$ is the channel coefficient between the $k$th interferer and the desired RX. $\delta_k(t)$ is a Bernoulli random variable that captures the traffic intensity and depends on the parameters $\mu$ and $l$, 
\begin{equation}
\label{eq.bernoul}
    \delta_k(t) = \left\{ \begin{array}{l}
1,~{\textrm{ if interferer $k$ is active depending on $\mu$ and $l$,}}\\
0,~{\textrm{ otherwise.}}
\end{array} \right. 
\end{equation}


The TX sends a short packet of $b$ bits with a target outage $\epsilon$ to the desired RX.
URLLC aims to simultaneously achieve high reliability and low latency, typically by meeting a reliability target under a specific latency constraint. 
Unlike conventional systems, retransmissions are impractical here to enhance reliability, not only due to the additional time overhead but also because control systems often require fresh information; instead, consecutive outages must be mitigated to maintain control stability.

\subsection{Link Quality Metrics}
The SINR at the desired RX location and timestep $t$ is $\rho_t=p_t|h_t|^2/(\sigma^2+I_t)$, where $\sigma^2$ is the noise power, and $I_t$ is the sum of interference power at the desired RX location, as defined by $I(t)$ in \eqref{eq.sumInterf}. In the finite blocklength regime, the outage probability, $\epsilon_t$, of a block with $b$ information bits transmitted in a blocklength of $m_t$ channel uses in an additive white Gaussian noise (AWGN) channel with SINR $\rho_t$ is given as \cite{polyanskiy}
\begin{equation}
\label{eq.FBL}
    \epsilon_t \approx Q\left(\frac{C(\rho_t)-\frac{b}{m_t}}{\sqrt{V(\rho_t)/m_t}}\right), 
\end{equation}
 where $C(\rho_t)=\log_2(1+\rho_t)$ is the Shannon capacity of AWGN channel, $V(\rho_t)=(\log_2(e))^2(1-(1+\rho_t)^{-2})$ is the channel dispersion, and $Q(.)$ is the Gaussian Q-function. 
Since the transmission occurs over mini-slots in URLLC, the channel coherence time covers several transmission intervals. Hence, the characteristics of our channel model remain the same in each realization and can be assumed as an AWGN channel.

To quantify performance, we consider three different link quality metrics based on the outage probability as follows.

\textbf{Reliability:}
Reliability refers to the probability that a transmission is successfully completed within a specified time frame \cite{MC_availability}, typically measured as the complement of block error ratio (BLER) or outage probability, i.e., $1-\epsilon_t$. This metric captures the ability of the communication system to deliver packets successfully in one-shot transmissions without requiring retransmissions, which is crucial for mission-critical applications like autonomous vehicles and industrial automation. To ensure reliability, the outage probability stated in \eqref{eq.FBL} is to be minimized under a certain application-specific threshold (i.e., $\epsilon_t \leq \varepsilon^{\mathrm{th}}$).

\textbf{Availability:}
Availability refers to the long-term operational readiness of the communication link, defined as the proportion of time during which the system can support reliable communication. Mathematically, it is expressed as the probability that the instantaneous block error rate satisfies the reliability threshold, i.e.,
\begin{equation}
\label{eq.avail}
   P_A = \operatorname{Pr}\left \{\epsilon_t \leq \varepsilon^{\mathrm{th}}\right \}= \lim_{T \to \infty} \frac{1}{T}\sum_{t=1}^T \mathbbm{1}(\epsilon_t \leq \varepsilon^{\mathrm{th}}) , 
\end{equation} 
where $\mathbbm{1}\mathcal(x)$ is the indicator function, which equals $1$ if $x$ is true and  $0$ otherwise. In contrast to reliability, which measures the success probability of a single transmission, availability captures the overall system uptime, factoring in persistent disruptions such as network failures, downtime, or scheduled maintenance in industrial environments \cite{missionReliability}. Ensuring high availability is essential for factory automation scenarios, where communication interruptions can lead to safety risks or production halts.

\textbf{Burst/Consecutive Outages:}
Burst/consecutive outages measure the occurrence of sequential transmission failures or outages over a certain time window. While individual outages may be tolerated occasionally in some applications, consecutive outages indicate a persistent degradation in link quality.
For example, a series of consecutive transmission failures could lead to severe interruptions in critical processes, such as industrial control systems or telemedicine. Monitoring and minimizing consecutive outages is important for ensuring both short-term reliability and the stability of communication links over time.
This parameter is widely used in network control system design, particularly in models based on deterministic packet losses \cite{dropout}.
We numerically assume an outage occurs when the outage probability exceeds the desired threshold, i.e., $\mathbbm{1}(\epsilon_t > \varepsilon^{\mathrm{th}})$. Hence, the number of consecutive outages at timestep $t$ is recursively defined as $C_{\epsilon,t} = (C_{\epsilon,t-1}+1)\mathbbm{1}(\epsilon_t > \varepsilon^{\mathrm{th}})$.

\subsection{Problem Formulation}
The main goal of this work is to adjust the allocated power and blocklength of the desired URLLC link under availability constraint to optimize two objectives: consecutive outages and resource efficiency. 
These two objectives are inherently conflicting. Reducing the transmit power and blocklength to increase resource efficiency also increases the outage probability.
The first objective function, $F_1$, reflects the consecutive outage probability and is defined as, 
\begin{equation}
\label{eq.F1}
F_1(\mathit{C_{\epsilon, t}})= \operatorname{Pr}\left \{\mathit{C_{\epsilon,t}} > L^{\mathrm{th}} \right \} = \lim_{T \to \infty} \frac{1}{T}\sum_{t=1}^T \mathbbm{1}({C_{\epsilon,t}} > L^{\mathrm{th}}),
\end{equation}
which indicates the probability that the number of consecutive outages exceeds a certain value $L^{\mathrm{th}}$.

At the same time, the system aims to minimize the consumed energy to enhance EE. This objective function, $F_2$, is defined as,
\begin{equation}
\label{eq.F2}
    F_2(\mathit{E_t})= \mathbb{E}\{ {\mathit{EE}} \} = \lim_{T \to \infty} \frac{1}{T} \sum_{t=1}^T \frac{b}{p_t  m_t},
\end{equation}
which represents the number of bits $b$ transmitted per unit of consumed energy according to the allocated power $p_t$ and blocklength $m_t$ at timestep $t$.
Additionally, optimal resource allocation becomes more challenging due to random channel conditions and interference power. 

Based on the objectives in \eqref{eq.F1} and \eqref{eq.F2}, the optimization problem in this work involves finding the optimal power and blocklength for all timesteps $t\in \{1,2,\cdots,T\}$ according to the following chance-constrained problem to minimize the objective function $F_1$ while maximizing the objective function $F_2$,
\begin{subequations}
\label{eq.opt}
\begin{align}
\mathop {\max }\limits_{\{p_t, m_t\}}& {\;\;  - F_1(\mathit{C_{\epsilon, t}}) + \omega F_2(\mathit{E_t})} \label{eq.a}\\
{\rm{ s.t.}}\quad& \operatorname{Pr}\left \{\epsilon_t \leq \varepsilon^{\mathrm{th}}\right \} \geq A^{\mathrm{th}},\label{eq.b}\\
&0 < {p_t}\leq P_{\mathrm{max}} ,\label{eq.c}\\
&{m_t}\leq M_{\rm{max}}, m_t \in \mathbb{Z}^+ .\label{eq.d}
\end{align}
\end{subequations}
The constraint in \eqref{eq.b} ensures both reliability in individual transmission $t$ and link availability in the long-term period $T$. It states that the outage probability should be lower than the target outage $\varepsilon^{\mathrm{th}}$ and at the same time the link availability should be greater than the target value $A^{\mathrm{th}}$. The constraint in  \eqref{eq.c} shows that the non-negative transmit power is limited by $P_{\rm{max}}$. Constraint  \eqref{eq.d}, which ensures the latency constraint, limits the allocated blocklength to a positive integer below $M_{\rm{max}}$. 
The weight $\omega \geq 0$ controls the importance of energy consumption and network performance.

The problem \eqref{eq.opt} is a mixed-integer optimization problem because blocklength $m_t$ would be an integer variable, while the power $p_t$ remains a continuous variable.
We first relax the integer constraint for blocklength $m_t$, i.e., $m_t \in \mathbb{R}^+$. The continuous blocklength solution is then deterministically projected to the feasible integer domain using an upward rounding operation, i.e., $m_t = \min(\lceil m_t \rceil, M_{\rm{max}})$, ensuring a valid blocklength.

Furthermore, the presence of stochastic elements like random channel conditions and interference power complicates finding the optimal solution. The probability-based objective function $F_1(\mathit{C_{\epsilon}})$ and constraint \eqref{eq.b} involve expectations over random variables, which generally result in non-linear and non-convex functions. Non-linear constraints and objectives, especially with probabilistic conditions, are known to increase computational complexity significantly \cite{meta_RL}. The presence of stochastic elements, combined with the non-linear probabilistic objective and constraints, renders the problem NP-hard. Additionally, the imposed bounds on power and blocklength further restrict the solution space, adding to the challenge of finding feasible solutions that satisfy all conditions, which overall makes the problem even more challenging. In such situations, RL excels at handling chance-constrained sequential decision-making under uncertainty and chance-constrained systems, where the agent learns a policy to optimize long-term cumulative rewards. In the next section, we solve problem \eqref{eq.opt} using DRL for power and blocklength allocation to consider the long-term benefits of allocating resources besides their immediate effects on the user’s performance.


\section{Deep Reinforcement Learning Approach}\label{sec.method}

\subsection{Markov Decision Process Formulation}
RL problems are formulated based on the MDP, which is the agent’s interaction with different states of the environment to maximize the expected long-term reward. Let $\mathcal{S}$ denote a set of possible states and $\mathcal{A}$ indicate the set of possible actions. In each timestep, the agent interacts with the environment, observes the current state $s_t \in \mathcal{S}$, takes action $a_t \in \mathcal{A}$ according to the current policy $\pi(s_t,a_t)$, where $\pi(s_t,a_t):\mathcal{S} \rightarrow \mathcal{A}$ is the probability of taking action $a_t$ given the current state $s_t$, receives an immediate reward $r_t$ and thereafter transitions to a next state $s_{t+1}$. This process is repeated, and in each timestep, the agent collects the tuple $\langle s_t, a_t, r_t, s_{t+1} \rangle$ as an experience. The objective of the RL-based system design is to find an optimal policy that maximizes the expected return. We design problem \eqref{eq.opt} as an MDP and define the state and action spaces and reward function as follows.

\textbf{State Space:}
The state space describes the environment in which the agent is interacting through different actions. We define the state at timestep $t$ as $s_t = \{
\varrho_{t-1}, C_{\epsilon,t-1}, P_{A,t-1}\} \in \mathcal{S}$, where $\varrho_{t-1} = |h_{t-1}|^2/(\sigma^2+I_{t-1})$ is the delayed channel quality, while $C_{\epsilon,t-1}$ and $P_{A,t-1}$ are the number of consecutive outages and the link availability up to the previous timestep, respectively.

\textbf{Action Space:}
The action space includes all possible actions that the agent can take. The action can change the state of the environment from the current state to the target state. In our problem, $a_t = (p_t,m_t) \in \mathcal{A}$ is the decision regarding the allocated power and blocklength at timestep $t$.

\textbf{Reward Function:}
The agent obtains the reward after taking action $a_t$ when the current state is $s_t$ and moves to the next state $s_{t+1}$. The reward function is defined by modifying the objective function and constraints in the optimization problem \eqref{eq.opt}. Because the learning agent is designed to maximize its reward, we incorporate two reliability penalty terms associated with consecutive-outage exceedance and availability violation as negative terms. The reward $r_{t}$ obtained by the RL agent at timestep $t$ is composed of reward elements for different objectives. Hence, the reward $r_t$ is defined as
\begin{equation}
\label{eq.rew}
    r_t= \omega \widehat{\mathit{EE}} - \max(0,C_{\epsilon} - L^{\mathrm{th}}) - \frac{\max(0, A^{\mathrm{th}}-P_A)}{A^{\mathrm{th}}},
\end{equation}
where $\widehat{\mathit{EE}}$ is the normalized EE with min-max normalization, the second term is the raw number of consecutive outages exceeding the threshold, and the third term is the normalized availability gap. The coefficient $\omega$ controls the contribution of energy efficiency, while reliability penalties are assigned fixed unit weights to prioritize the satisfaction of reliability constraints, with longer outage bursts incurring increasingly severe penalties.

\subsection{SAC-Based Resource Allocation Algorithm}

\begin{figure}[!t]
    \centering           \includegraphics[width=1\columnwidth,clip]{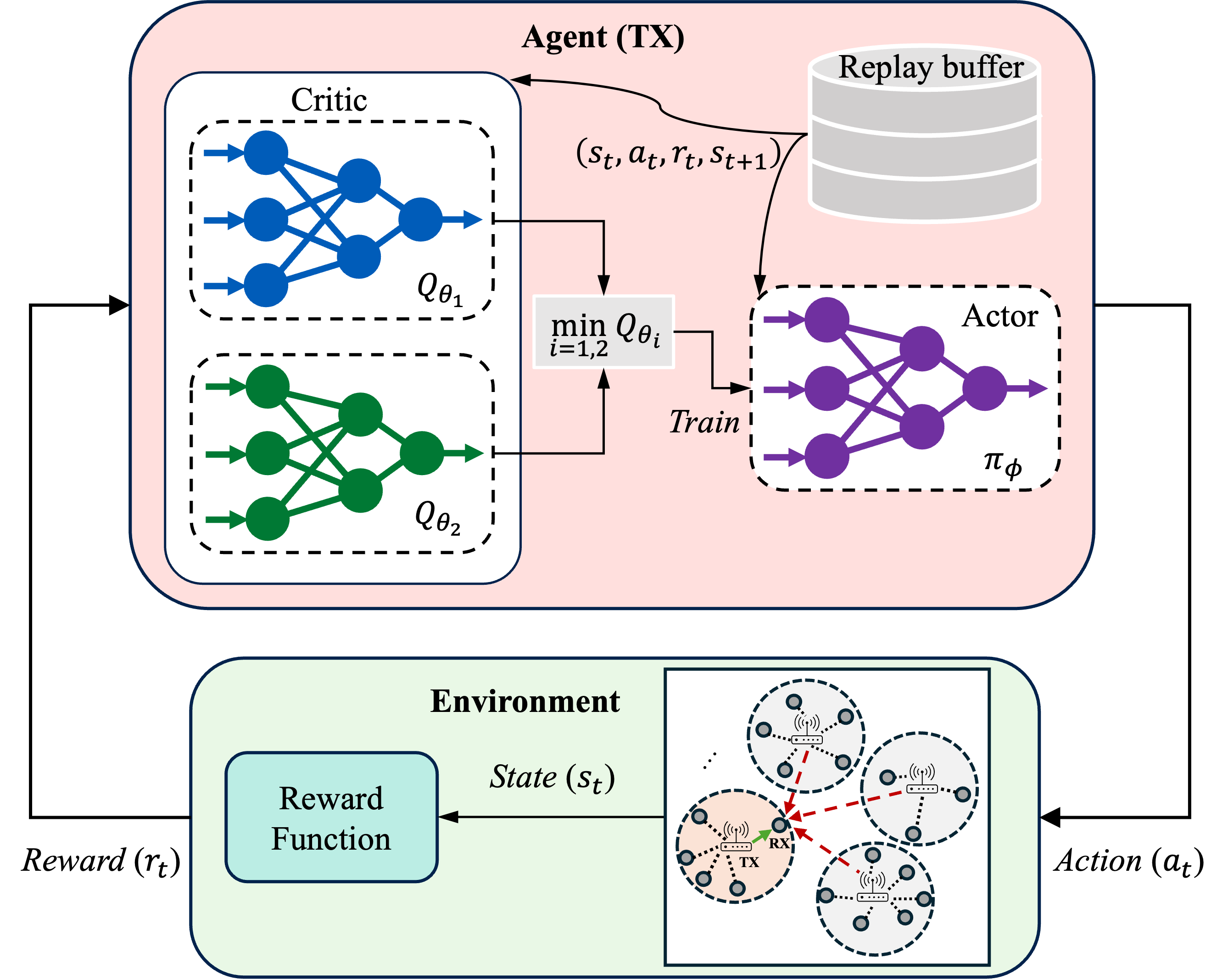}
    \vspace{-2mm}
    \caption{Structure of the proposed SAC-based algorithm for radio resource allocation.
            }            
    \label{fig.model}
    \vspace{-1.6em}
\end{figure} 

Soft Actor-Critic (SAC) is an advanced RL technique that belongs to the class of off-policy, model-free algorithms and combines elements from both actor-critic methods and entropy-regularized RL. By explicitly maximizing the entropy of the policy, SAC optimizes not only for high rewards but also for high stochasticity, which helps the agent avoid suboptimal solutions in dynamic and uncertain environments like our problem. This entropy maximization promotes a balance between exploration and exploitation, enabling the agent to explore a more diverse set of power and blocklength actions and environments effectively \cite{sac}. 
In SAC, the policy is stochastic, meaning it outputs a distribution over resource allocation actions, not just a single deterministic action. These features make SAC particularly suitable for balancing reliability and EE in URLLC scenarios. As shown in Fig. \ref{fig.model}, the algorithm uses two main networks:
\begin{itemize}
    \item A policy network (actor), \(\pi_{\phi}(a_t | s_t)\), that learns a distribution over actions.
    \item Two Q-value networks (critics), \( Q_{\theta_1}(s_t, a_t) \) and \( Q_{\theta_2}(s_t, a_t) \), that estimate the expected return of a given state-action pair \((s_t, a_t)\). 
\end{itemize}

Having two Q-networks helps stabilize learning by reducing overestimation bias in Q-value predictions, offering an advantage over methods like DDPG.
The critics' parameters \(\theta_1\) and \(\theta_2\) are updated using a soft Q-learning update to minimize the Bellman error with entropy regularization.
For a stochastic policy \(\pi_\phi(a_t | s_t)\), the entropy \( \mathcal{H} \) is defined as
\begin{equation}
    \mathcal{H}(\pi_\phi) = - \mathbb{E}_{a_t \sim \pi_\phi} \left[ \log \pi_\phi(a_t | s_t) \right],
\end{equation}
where \( \pi_\phi(a_t | s_t) \) is the probability of taking action \( a_t \) in state \( s_t \) under the policy parameterized by \(\phi\).
The soft Q-value function in SAC for policy $\pi_{\phi}$ being followed is given as
\begin{equation}
\begin{split}
    &Q^{\pi_{\phi}}(s, a) =\\ &\mathbb{E}_{\pi_{\phi}} \left[ \sum_{t=0}^\infty \gamma^t \left( r_t - \alpha \log \pi_{\phi}(a_t | s_t) \right) \mid s_0 = s, a_0 = a \right],
\end{split}
\end{equation}
where \( r_t(s_t, a_t) \) is the reward received after taking action \( a_t \) in state \( s_t \), and \( \gamma \) is the discount factor. The entropy term, \(\alpha \log \pi_{\phi}(a_t | s_t)\), promotes exploration by encouraging the policy to stay stochastic, and \( \alpha \) is the temperature parameter, controlling the trade-off between reward and entropy maximization. The soft Q-value target, which incorporates both the reward and the entropy of the action for each state-action pair is defined as
\begin{equation}
\label{eq.Qvalue}
\begin{split}
&y_t = r_t +\\ 
&\gamma \mathbb{E}_{s_{t+1} \sim P} \!\!\left[\! \min_{i=1,2} Q_{\theta_i}\!(s_{t+1}, a_{t+1}) \!-\! \alpha \log \pi_{\phi}(a_{t+1} | s_{t+1})\! \right]\!,
\end{split}
\end{equation}
where \( a_{t+1} \sim \pi_{\phi}(a_{t+1} | s_{t+1}) \) is the action sampled from the policy at the next state \( s_{t+1} \). SAC can also reduce overestimation by using the minimum Q-value between the two critics, i.e., \(\min_{i=1,2} Q_{\theta_i}(s_{t+1}, a_{t+1})\).

Target value $y_t$ is then used to compute the critic loss as
\begin{equation}
\label{eq.critic}
    L_Q(\theta_i) = \mathbb{E}_{(s_t, a_t, r_t, s_{t+1}) \sim \mathcal{D}} \left[ \left( Q_{\theta_i}(s_t, a_t) - y_t \right)^2 \right],
\end{equation}
where \(\mathcal{D}\) is the replay buffer containing past experiences.
The policy is updated to maximize the expected Q-value while also maximizing entropy. The optimal policy $\pi^*$ is learned by minimizing the following loss function
\begin{equation}
\label{eq.actor}
L_{\pi}(\phi)\!=\!\mathbb{E}_{s \sim \mathcal{D}, a_t \sim \pi_{\phi}} \!\!\left[\alpha \log \pi_{\phi}(a_t | s_t) - \min_{i=1,2} Q_{\theta_i}(s_t, a_t) \right].
\end{equation}
The temperature parameter \(\alpha\) can be adjusted automatically to reach a target entropy level, improving stability and adaptability in different environments.
This approach allows SAC to maintain a high level of performance while adapting to the ever-changing conditions of complex systems. 

In Algorithm \ref{alg.RL}, we provide the proposed SAC-based RL design to solve our joint optimization problem.
At first, the actor and critic network parameters, i.e., \(\phi \), \(\theta_1 \), \(\theta_2 \), are randomly initialized for training (line 1 of the algorithm). The policy is updated via SAC in lines 2-12. First, the agent observes the current state $s_t$ and selects an action \(a_t \sim \pi_{\phi}(a_t|s_t)\) based on the policy (line 3). Then, the action $a_t$ is executed, and the next state $s_{t+1}$ is observed (line 4).
The training loop continues until the policy and Q-values converge, optimizing the trade-offs between reliability and resource efficiency. 
The trained model is finally evaluated on the unseen test environment.

\begin{algorithm}
\DontPrintSemicolon
 \textbf{Initialization}: Initialize policy parameter \(\phi\), Q-value parameters \(\theta_1 \) and \(\theta_2 \), replay buffer \(\mathcal{D}\), batch size $B$. Set target parameters equal to main parameters $\theta_{\mathrm{target},i} \leftarrow \theta_i$ for $i = 1,2$. \\
\Repeat{convergence}
{Observe state $s_t$ and select action \(a_t \sim \pi_{\phi}(.|s_t)\).\;
Execute $a_t$ in the environment and observe the next state $s_{t+1}$.\;
Compute outage probability $\epsilon_t$, link availability $P_A$, number of consecutive outages $C_{\epsilon,t}$, and energy efficiency $\mathit{EE_t}$, using equations \eqref{eq.FBL} to \eqref{eq.F2}, then calculate immediate reward $r_t$ via \eqref{eq.rew}. \;
Store $(s_t, a_t, r_t, s_{t+1})$ in replay buffer \(\mathcal{D}\). \;
Randomly sample a batch $B$ from the replay buffer. \;
Compute targets for the Q-values using \eqref{eq.Qvalue}. \;
Update Q-values by one gradient descent step using \eqref{eq.critic} over batch $B$ for $i = 1,2$. \\
Update policy by one gradient descent step using \eqref{eq.actor} over batch $B$. \\
Update target networks with $\theta_{\mathrm{target},i} \leftarrow \nu \theta_{\mathrm{target},i} + (1 - \nu) \theta_i$ for $i =1,2$.
}
\caption{Soft Actor-Critic (SAC) for Mitigating Consecutive Outages and Minimizing Resource Usage}
\label{alg.RL}
\end{algorithm}


\section{Numerical Results}\label{sec.res}
This section analyzes the empirical results through the key performance metrics, i.e., link availability, consecutive outages, and energy consumption. We implemented the proposed architecture and baseline models in Python using PyTorch and Stable Baselines-3 \cite{SB3} for the DRL algorithms. All experiments were conducted on a high-performance computing (HPC) system equipped with an HPE Cray XD685 server featuring two AMD EPYC 9655 processors (96 cores each, 2.6 GHz; 384 vCPUs in total), 2,304 GB of DDR5-6400 memory, and eight NVIDIA Blackwell B200-SMX6 GPUs, each with 192 GB of memory.
The model configurations and key hyperparameters are detailed in Table \ref{tab:rl_configs}.

Each DRL algorithm was trained over 20 independent trials using controlled random seeds.
During training, the policy was periodically evaluated using an evaluation callback. The model was evaluated every 2000 timesteps over two episodes, and the model achieving the highest mean evaluation reward was saved. To avoid unnecessary training after convergence, an early-stopping criterion was applied: after 10 evaluations, training was stopped if no improvement in the best mean reward was observed for 50 consecutive evaluations. This procedure ensures efficient training while preserving the best-performing policy for subsequent performance evaluation.

The trained models were then evaluated on the unseen test set using deterministic policy execution. Training, evaluation, and testing were performed on separate channel/interference traces. Training episodes used randomized starting points within the training trace, while evaluation and testing were performed on fixed deterministic windows to ensure fair comparison across trials. The best model in each trial was selected based on the mean evaluation reward and saved for subsequent testing. 
The testing phase consists of $10^6$ timesteps, and the final results are obtained by averaging over 20 trials.

The performance of the proposed SAC-based solution from Section \ref{sec.method} is numerically evaluated against the following baselines:
\begin{itemize}
    \item \textit{Deep Reinforcement Learning (DRL) algorithms:} The proposed method is compared with other DRL algorithms \cite{DRL_survey} such as deep deterministic policy gradient (DDPG), twin Delayed DDPG (TD3), Advantage Actor-Critic (A2C), and proximal policy optimization (PPO).
    \item \textit{Random Assignment (RA):} The resources are allocated to the user arbitrarily based on independent and uniform distributions in the feasible range of transmit SNR (in dB) and the blocklength, serving as a baseline to evaluate performance gains without any optimization effort.
    \item \textit{Maximum Resources (MR):} Always, the maximum amount of power and blocklength is allocated to the user. Although it is practically inefficient, we consider it as an indicator of the best performance bound on reliability.
\end{itemize}

\begin{table*}[t]
\centering
\caption{Model Configurations and Key Hyperparameters for SAC, DDPG, TD3, A2C, and PPO.}
\label{tab:rl_configs}
\vspace{-7pt}
\resizebox{\textwidth}{!}{%
\begin{tabular}{|l|l|l|l|l|l|l|l|l|}
\hline
\textbf{Algorithm} & \textbf{Actor Network}          & \textbf{Critic Network}          & \textbf{Activation Function} & \textbf{Learning Rate} & \textbf{Optimizer} & \textbf{Replay Buffer Size} & \textbf{Other Parameters} & \textbf{Shared/Separate} \\ 
\hline
\hline
\textbf{SAC}       & 2 layers, 256 neurons each      & 2 layers, 256 neurons each       & ReLU                        & 3e-4                  & Adam              & 1,000,000                  & 0.2 $^1$                        & Separate                 \\ 
\textbf{DDPG}      & 2 layers, 400 and 300 neurons   & 2 layers, 400 and 300 neurons    & ReLU                        & 1e-3                  & Adam              & 1,000,000                  & $\mathcal{N}(0,0.01)$    $^2$                     & Separate                 \\ 
\textbf{TD3}       & 2 layers, 400 and 300 neurons   & 2 layers, 400 and 300 neurons    & ReLU                        & 1e-3                  & Adam              & 1,000,000                  & $\mathcal{N}(0,0.01)$   $^2$                       & Separate                 \\ 
\textbf{A2C}       & 2 layers, 64 neurons each       & Shared with actor                & ReLU                        & 7e-4                  & RMSProp           & N/A                        & 5   $^3$                      & Shared                   \\ 
\textbf{PPO}       & 2 layers, 64 neurons each       & Shared with actor                & ReLU                        & 3e-4                  & Adam              & N/A                        & 2048       $^3$                  & Shared                   \\ \hline

\multicolumn{9}{p{502pt}}{$^1$ Entropy Coefficient \hspace{0.5cm}
$^2$ Action Noise \hspace{0.5cm}
$^3$ Rollout Length }\\

\multicolumn{9}{p{550pt}}{\textbf{Notes:} For A2C and PPO, the actor and critic networks share the same layers. Replay buffer size is applicable only for off-policy algorithms (DDPG, TD3, SAC).}

\end{tabular}%
}



\vspace{-0.5cm}
\end{table*}

\subsection{Experiment Setup}
Table \ref{tab.setup} summarizes the simulation settings for the performance evaluation. The maximum value of transmit SNR represented by \(\zeta\), i.e. \(\zeta = p/\sigma^2\), is assumed \(\zeta_{\rm{max}} = 20\)\,dB, and the maximum blocklength is assumed \(M_{\mathrm{max}}=10^3\).  The mean interference-to-noise ratio (INR), i.e., $I/\sigma^2$, is modeled as a random variable uniformly distributed within the range of \([-10, 5]\)\,dB. 
The target outage probability \(\varepsilon^{\mathrm{th}}=10^{-5}\), and the desired link availability \(A^{\mathrm{th}}=0.99\) in a scenario involving five interfering subnetworks with activation factor $\mu = 1$. Here we assume $L^{\mathrm{th}} =5$.
The weight of energy efficiency is set as $\omega = 1$. 

\begin{table}[!t]
\centering
\caption{Simulation Parameters}
\vspace{-12pt}
\label{tab.setup}
\begin{center}
\begin{tabular}{ |l|c|l| } 
 \hline
\textbf{Parameter} & \textbf{Symbol} & \textbf{Value}\\
 \hline
 \hline
 Transmit SNR range & $\zeta$ & $[-20,20]$\,dB\\
 Blocklength range & $m$ & $[1,10^3]$ \\
 Mean INR range & & $\mathcal{U}[-10,5]$\,dB \\
 Packet size & $b$ & $50$ bits \\ 
 Number of interfering subnetworks & $N$ & $5$ \\ 
 Activation factor & $\mu$ & $1$ \\ 
 Message duration & $l$ & $10$ mini-slots \\ 
 Target outage probability & $\varepsilon^{\mathrm{th}}$ & $10^{-5}$ \\
 Link availability & $A^{\mathrm{th}}$ & $0.99$ \\
 Consecutive outage threshold & $L^{\mathrm{th}}$ & $5$ \\
 Weight of energy efficiency & $\omega$ & $1$ \\
 \hline
 \end{tabular}
\end{center}
\vspace{-0.5cm}
\end{table}

\subsection{Comparison of Convergence Performance}\label{subsec.conv}
\begin{figure}[t]
    \centering
        \includegraphics[width=1\columnwidth,trim={0cm 0.2cm 0cm 0cm},clip]{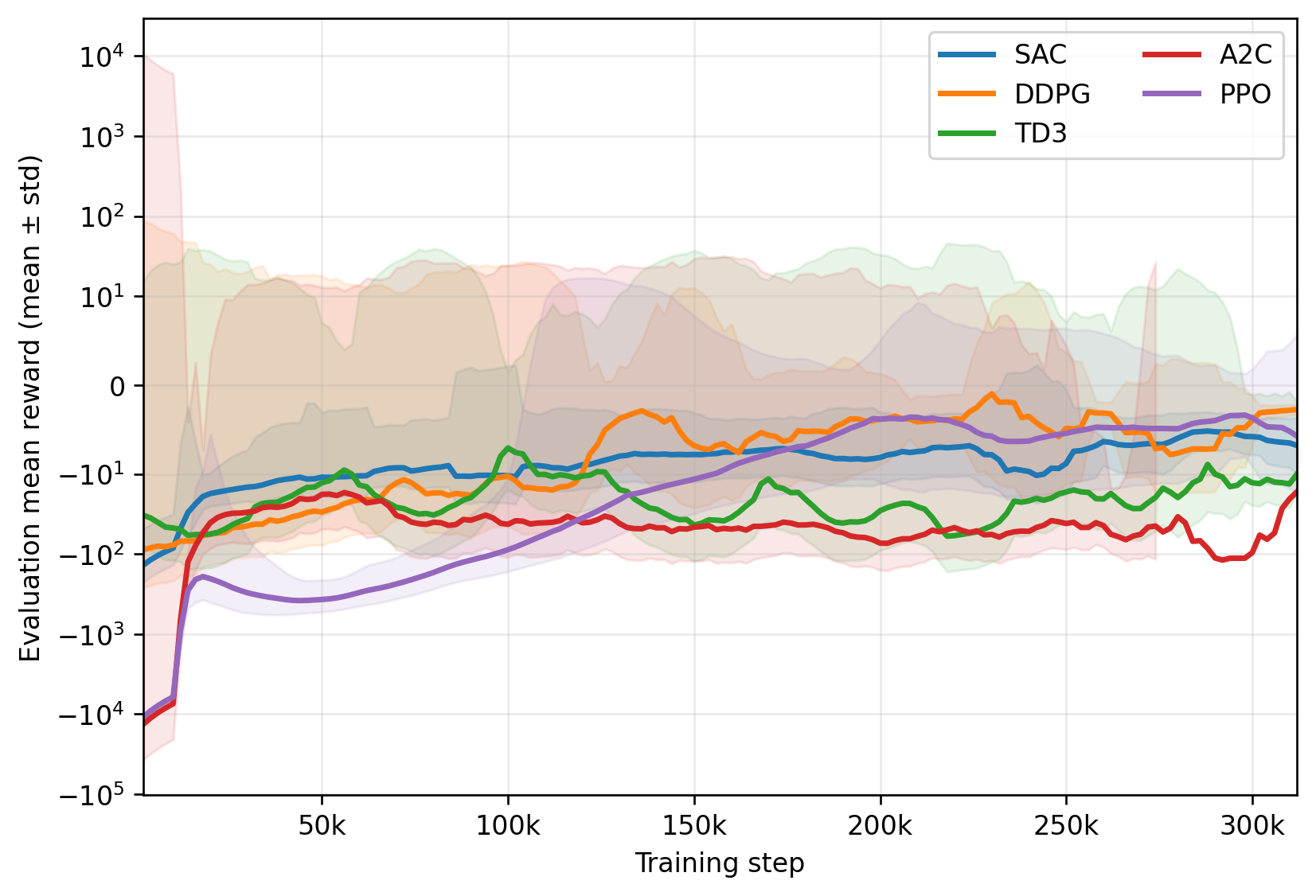}
            \vspace{-12pt}
            \caption{Training results of different DRL algorithms in terms of evaluation mean reward per training timestep.}
            \label{fig.reward}
            \vspace{-13pt}
\end{figure}

Fig. \ref{fig.reward} compares the convergence behavior of SAC, DDPG, TD3, A2C, and PPO over approximately $3.3\times10^5$ training steps. Each curve represents the mean evaluation reward averaged over 20 independent trials, while the shaded region indicates the variability across trials. Because each episode contains 1000 environment steps, the horizontal axis corresponds to roughly 330 training episodes. The symmetric logarithmic scale is used to display both the large negative rewards observed during early training and the smaller reward differences near convergence.

The algorithms exhibit different learning dynamics. SAC improves rapidly during the initial stage, increasing from approximately $-10^2$ to around -10 after approximately 20k steps, and subsequently follows a smooth and stable trajectory, although its final mean reward remains negative around -4. PPO begins with a substantially lower reward but improves consistently throughout training and eventually achieves a positive mean reward with comparatively moderate variability. This behavior indicates slower initial learning than SAC but better asymptotic performance.

DDPG and TD3 show less stable convergence. Both improve gradually after around 100k steps. DDPG achieves some of the highest positive rewards during the later stages of training. However, its mean reward fluctuates considerably, and the wide confidence region indicates strong sensitivity to trial initialization, exploration noise, or stochastic training effects. TD3 initially improves but subsequently exhibits repeated oscillations and abrupt performance changes, and its final performance remains negative.

Like PPO, A2C recovers from very low initial rewards; however, its performance begins to deteriorate after approximately 70k training steps. Throughout training, A2C achieves the lowest mean reward among the evaluated algorithms and shows oscillations toward the end of training. 
Therefore, convergence should not be assessed solely by whether the mean reward approaches zero; the final reward, learning speed, and inter-trial variability must also be considered when comparing the algorithms.
In the following, we present the average of the testing results over 20 trained models for each DRL algorithm in a new environment with random channel gains and interference power.

\subsection{Performance in Terms of Energy Consumption}\label{subsec.energy}
\begin{figure}[!t]
    \centering
  \subfloat[CDF of consumed resources.]{%
    \includegraphics[width=1\linewidth]{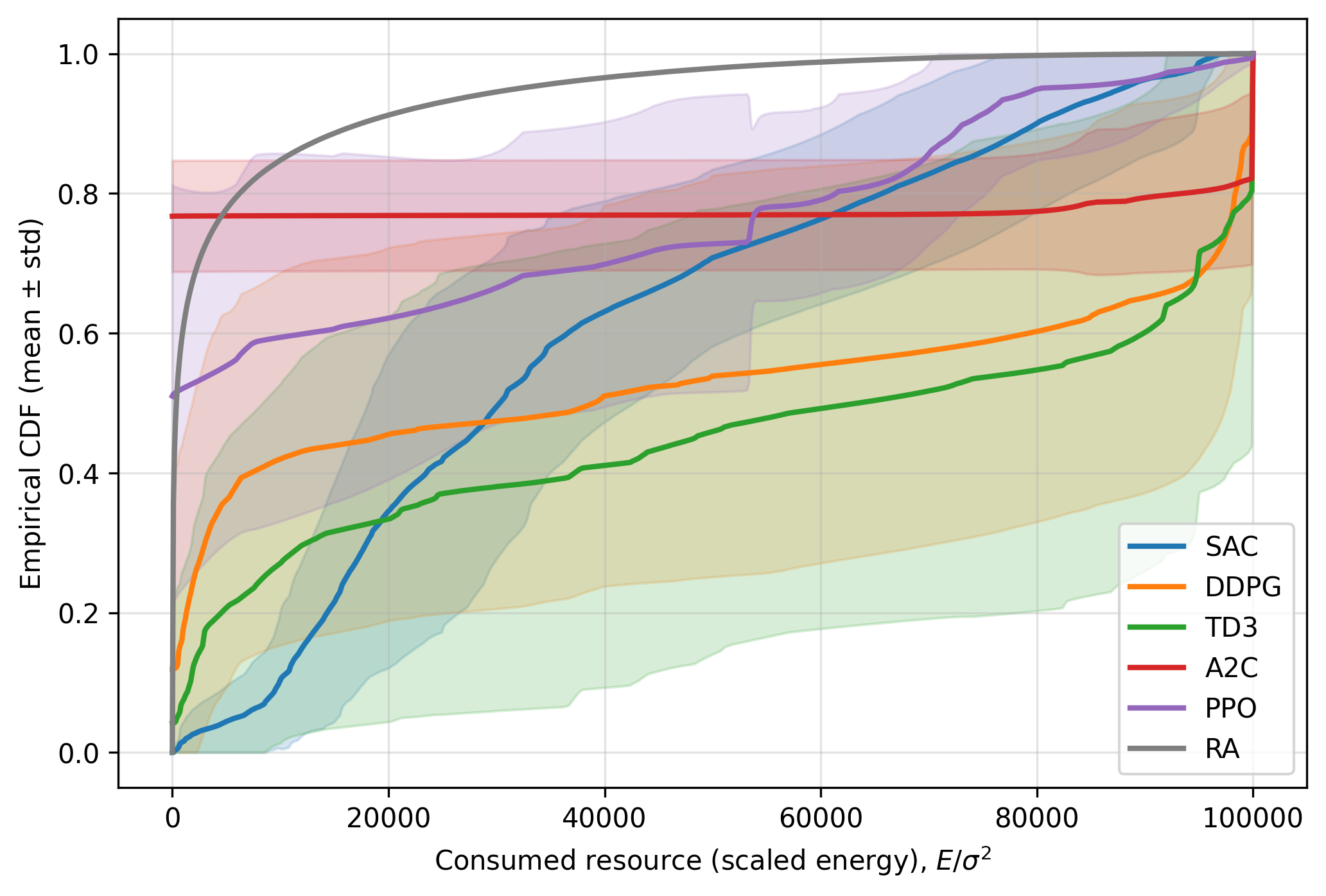}\label{figa}}\\
    \vspace{-10pt}
  \subfloat[Mean consumed resources.]{%
    \includegraphics[width=1\linewidth]{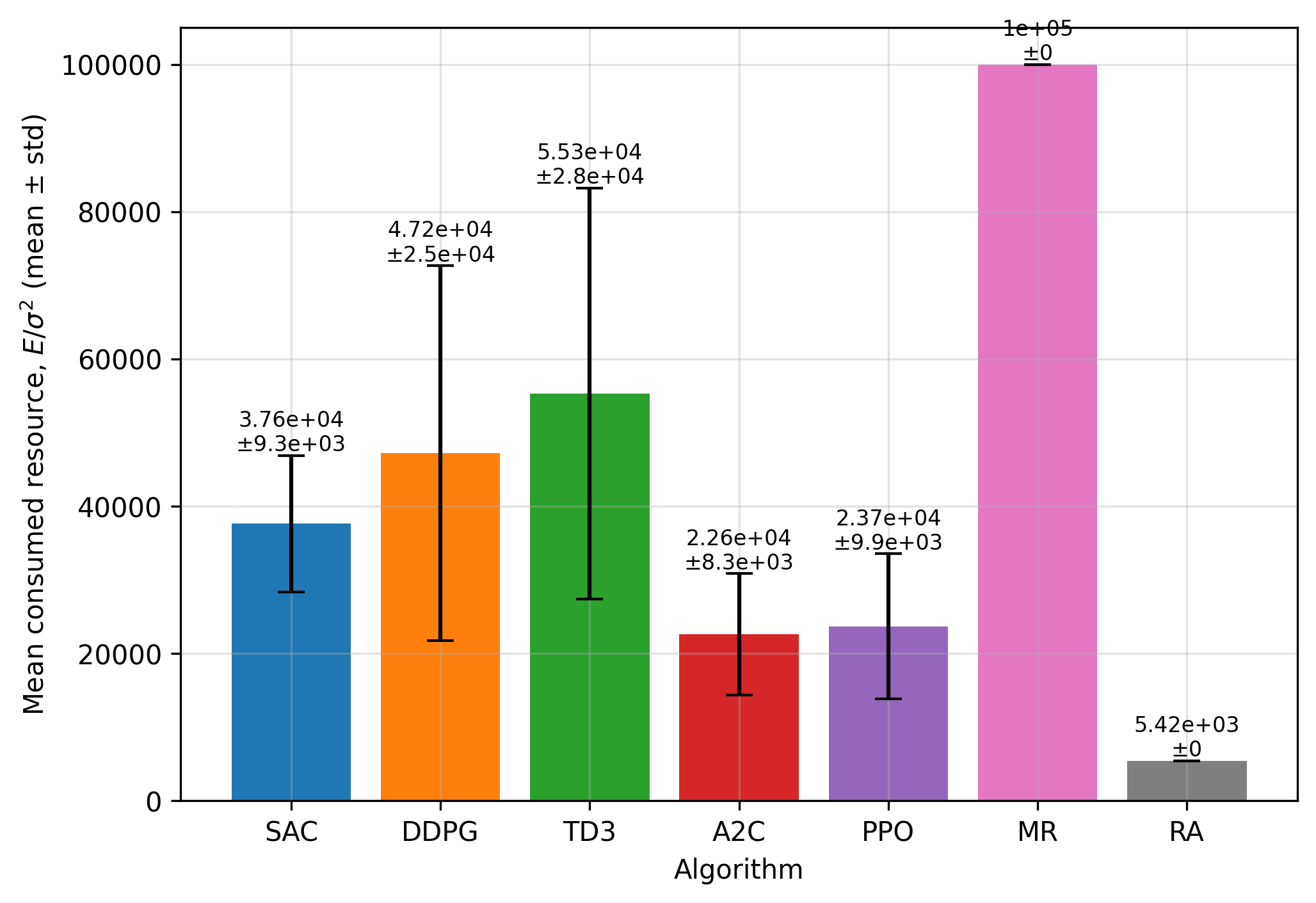}\label{figb}} 
  \caption{Comparison of resource consumption of different RRM schemes.}
  \label{fig.res}
  \vspace{-10pt}
\end{figure}

The CDF and mean value of the consumed resources, expressed as the scaled energy consumption, i.e., 
\(\zeta m = \frac{p \cdot m}{\sigma^2} = \frac{E}{\sigma^2}\), are presented in Fig.~\ref{fig.res}(a) and Fig.~\ref{fig.res}(b), respectively, for different RRM schemes. The maximum possible resource consumption is \(10^5\), corresponding to the MR scheme, which allocates the maximum transmit power and blocklength at all timesteps. As expected, MR achieves the highest resource consumption and serves as an upper-bound reliability benchmark.
The RA scheme with the CDF shifted to the upper-left consumes the lowest average resources, approximately \(5.42 \times 10^3\), but this low resource usage is obtained without reliability-aware optimization. 

Among the DRL-based schemes, A2C and PPO exhibit the lowest mean resource consumption, using approximately \(2.26 \times 10^4\) and \(2.37 \times 10^4\), respectively, corresponding to roughly \(22\%\)--\(23\%\) of the maximum resources. Their CDF curves show that these algorithms favor two strategies: first, minimum resource with an average probability of around 0.75 or 0.5 to increase EE, and second, maximum resource allocation to achieve reliability. It shows that these algorithms fail to find a balanced strategy satisfying both objectives simultaneously. SAC follows with an average consumption of approximately \(3.76 \times 10^4\), corresponding to about \(37\%\) of the maximum resource budget. SAC can effectively achieve a balanced strategy in diverse resource allocation among all algorithms. This is because of SAC's stochastic policy, which guarantees that every useful set of resource allocation actions will not be ignored. Finally, DDPG and TD3 are the least resource-efficient among the DRL schemes, consuming approximately \(4.72 \times 10^4\) and  \(5.53 \times 10^4\), or about \(47\%\)--\(55\%\) of the maximum resources, respectively, with considerable variability across trials. Therefore, resource consumption alone is not sufficient to determine the best-performing algorithm; it must be interpreted jointly with the reliability metrics, including link availability and consecutive-outage performance.

\subsection{Performance Analysis of Consecutive Outages}\label{subsec.consec}
\begin{figure}[!t]
    \centering
  \subfloat[CDF of consecutive outages.]{%
    \includegraphics[width=1\linewidth]{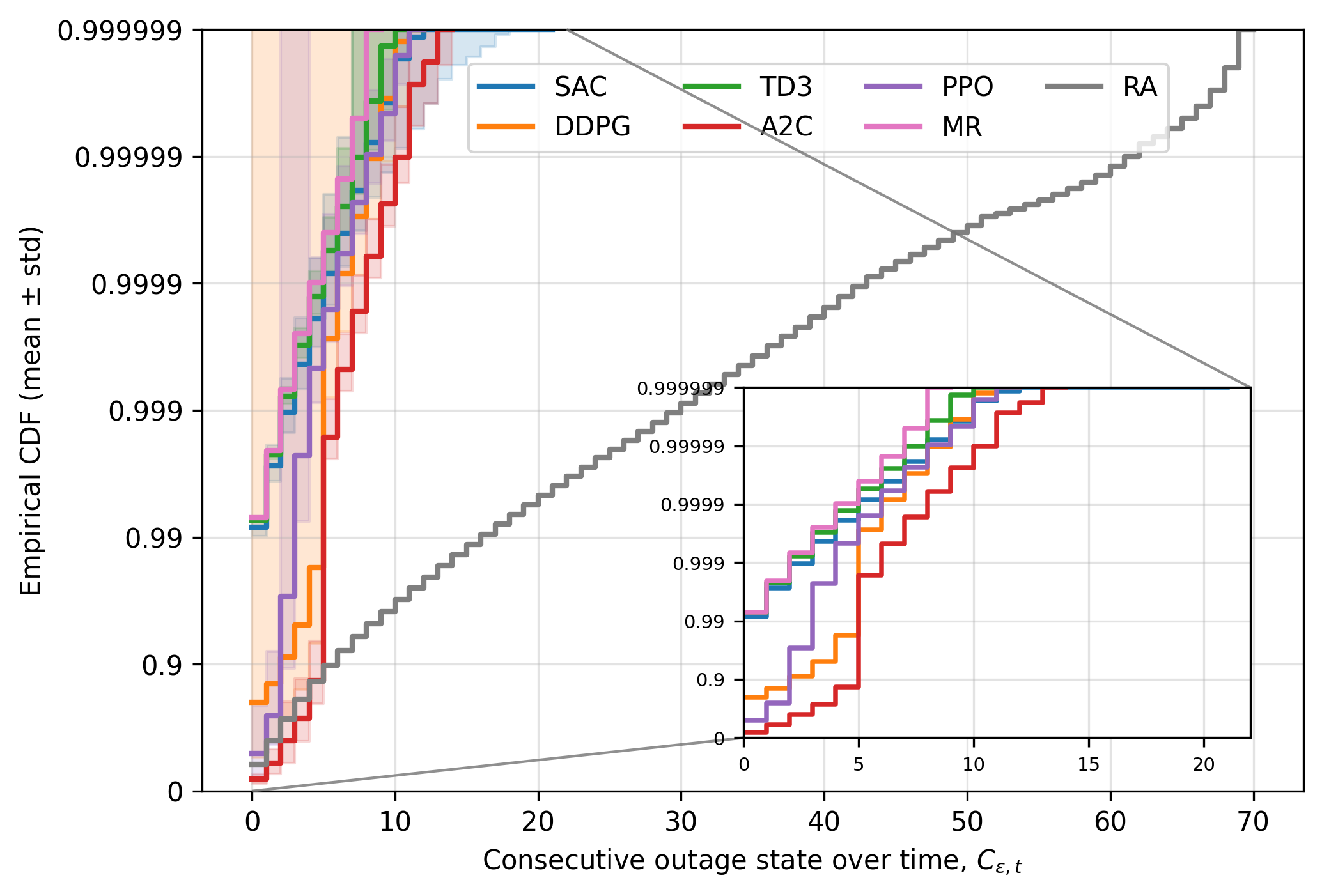}\label{figa}}\\
    \vspace{-10pt}
  \subfloat[Consecutive outage exceedance probability from threshold 5.]{%
    \includegraphics[width=1\linewidth]{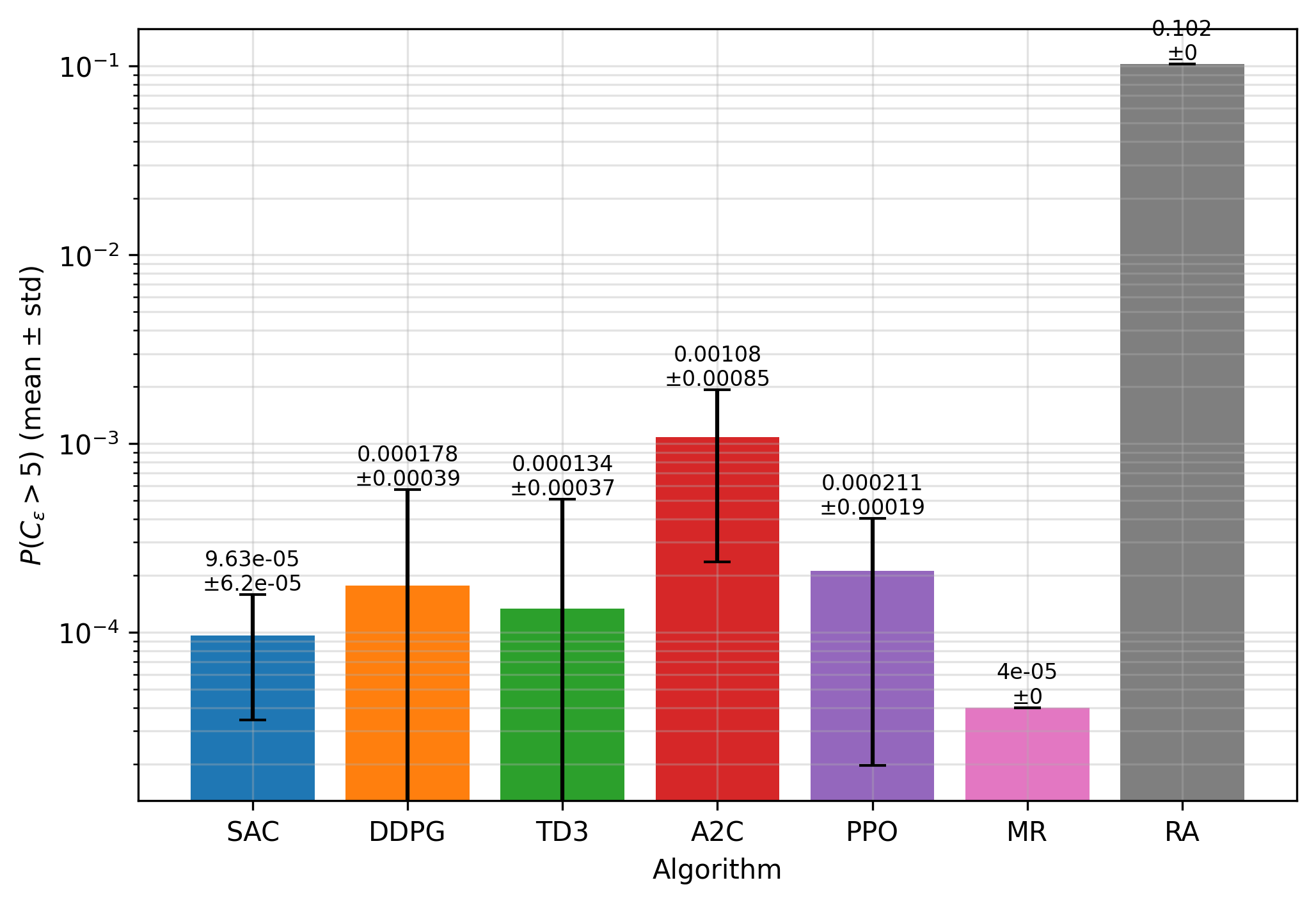}\label{figb}}
  \caption{Comparison of consecutive outage performance of different RRM schemes.}
  \label{fig.cons}
  \vspace{-10pt}
\end{figure}

Fig.~\ref{fig.cons} compares the consecutive-outage performance of the considered RRM schemes. Fig.~\ref{fig.cons}(a) shows the CDF of the consecutive-outage state \(C_{\epsilon,t}\) over time, while Fig.~\ref{fig.cons}(b) reports the exceedance probability \(\operatorname{Pr}\{C_{\epsilon,t}>5\}\). A steeper and more left-shifted CDF indicates that the algorithm experiences shorter outage bursts and therefore provides better reliability.

The MR scheme provides the best reliability benchmark, with the lowest consecutive-outage exceedance probability of approximately \(4\times 10^{-5}\). However, as discussed earlier, MR achieves this performance by always allocating the maximum resources and is therefore impractical for resource-efficient operation. Among the DRL-based schemes, SAC and TD3 achieve the best reliability performance, with exceedance probabilities of approximately \(9.63\times 10^{-5}\) and \(1.34\times 10^{-4}\), respectively. Their CDF curves are close to that of MR in the high-reliability region, indicating their ability to effectively suppress long consecutive-outage events.

PPO and DDPG also provide reliable performance, with exceedance probabilities of approximately \(1.78\times 10^{-4}\) and \(2.11\times 10^{-4}\), respectively, but they remain slightly behind SAC and TD3 in terms of tail reliability. The A2C algorithm shows weaker reliability, with an exceedance probability of approximately \(1.08\times 10^{-3}\), indicating that it is less suitable for stringent consecutive-outage constraints. Finally, RA performs the worst among all schemes, with a significantly larger tail and an exceedance probability of approximately \(0.102\), confirming that random resource allocation is unsuitable for ultra-reliable scenarios.

\subsection{Link Availability Performance Analysis}\label{subsec.avail}
\begin{figure}[t]
    \centering
    \includegraphics[width=1\columnwidth,trim={0cm 0.2cm 0cm 0cm},clip]{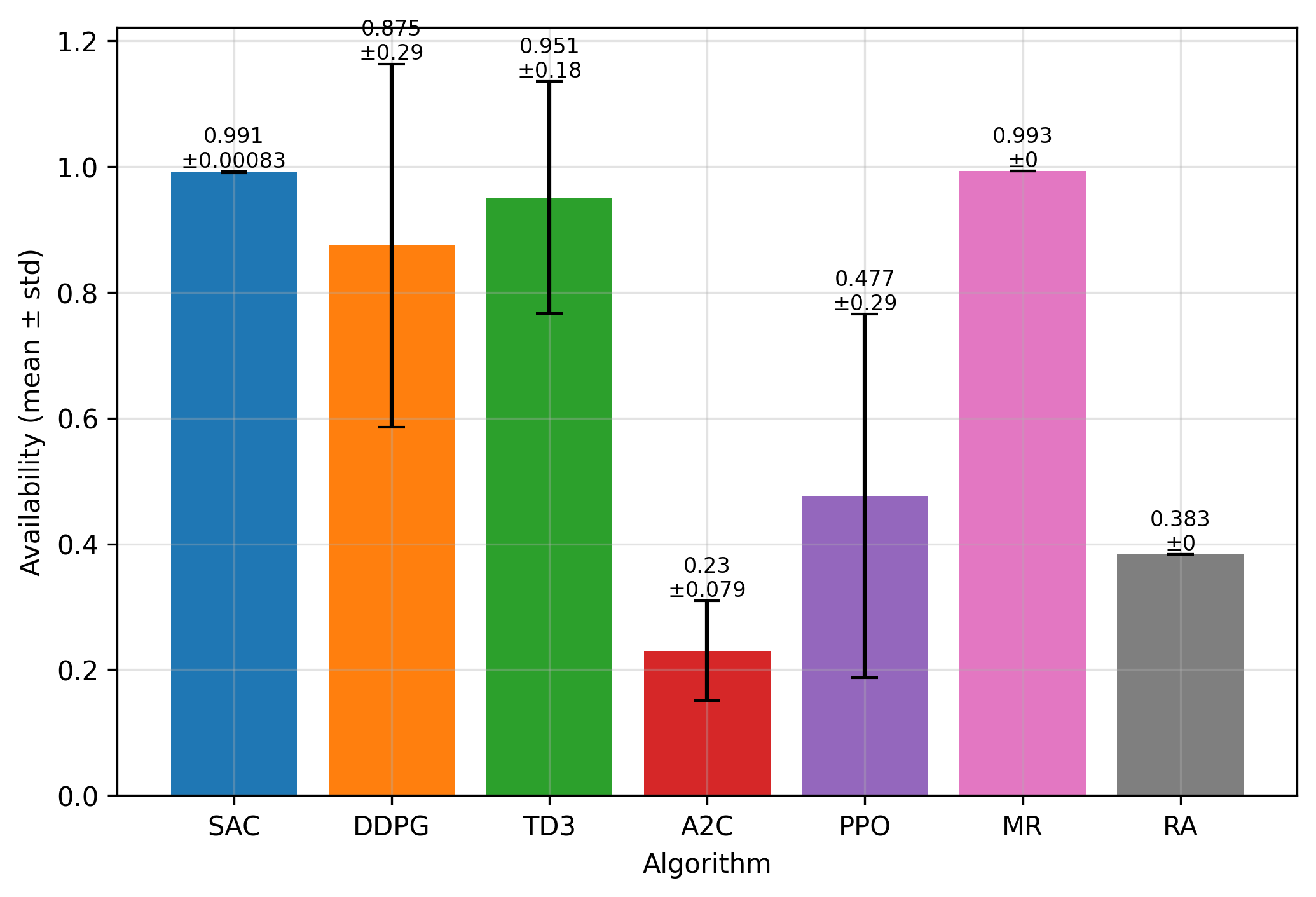}
            \vspace{-12pt}
            \caption{Link availability performance of different RRM schemes.}
            \label{fig.avail}
            \vspace{-10pt}
\end{figure}

Fig.~\ref{fig.avail} shows the link availability performance of different algorithms. The objective is to check whether the proposed algorithms can satisfy the constraint considered in \eqref{eq.b} or not. The assumed threshold for link availability is \(A^{\mathrm{th}} = 0.99\), selected according to the best achievable limit with the maximum allocation of resources.
MR represents the upper limit of the achievable link availability at 0.993.
Among DRL algorithms, only SAC can meet the threshold criteria with very low variance, suggesting that it maintains reliable performance and is a feasible solution for the considered scenario.
The TD3 and DDPG algorithms, however, have a mean value below the threshold with a large variance that can achieve the constraint in some trials, making them less reliable than the SAC algorithm. 
On the other hand, the other DRL algorithms (i.e., A2C and PPO) and the RA method also fail to meet the availability criterion. Their mean values are far from the threshold, indicating that these algorithms lack the acceptable link availability level in a significant proportion of cases. This makes A2C, PPO, and RA unsuitable for scenarios demanding high reliability.

\subsection{Comparison of Computational Complexity }
Table \ref{tab.time} compares the average per-step inference time of the considered algorithms when evaluated on a CPU. SAC has the highest inference time ($\sim$0.16\,ms). This is due to its stochastic nature (sampling-based optimization), soft policy updates, and use of entropy regularization, which involve more complex computations. 
DDPG ($\sim$0.116\,ms) and TD3 ($\sim$0.124\,ms) have comparable inference times, lower than SAC. These algorithms are deterministic and rely on actor-critic frameworks, which are computationally lighter compared to SAC due to the absence of entropy regularization. 
A2C ($\sim$0.071\,ms) and PPO ($\sim$0.067\,ms) are the fastest algorithms. Their efficiency stems from their simpler on-policy nature, which avoids storing and sampling large replay buffers.  

Based on the analyses in Fig.~\ref{fig.res}, Fig.~\ref{fig.cons}, and Fig.~\ref{fig.avail}, the proposed SAC-based method shows the best performance in reliability (captured by link availability and consecutive outage metrics) and resource efficiency, making it a strong candidate for real-world deployments where both are critical. However, its computational cost is slightly higher than the other algorithms. The need for training across multiple trials with varying reward weights requires the agent to learn from scratch each time. To mitigate this overhead, a pretraining strategy can be adopted, where a base SAC model is initially trained with a neutral reward configuration and subsequently fine-tuned for specific reliability-energy trade-offs. This not only reduces convergence time in later trials but also improves sample efficiency and learning stability.

\begin{table}[!t]
\centering
\caption{Mean Inference Time Per Step}
\vspace{-10pt}
\label{tab.time}
\begin{center}
\begin{tabular}{ |l|c|c| } 
 \hline
\textbf{Algorithm} & \textbf{mean$\pm$std (ms)}\\
\hline
\hline
 SAC & $0.16\pm 0.0022$\\
 DDPG & $0.116\pm0.0013$\\
 TD3 & $0.124\pm0.0009$\\
 A2C & $0.071\pm0.0027$\\ 
 PPO & $0.067\pm0.0012$\\ 
 \hline
 \end{tabular}
\end{center}
\vspace{-0.5cm}
\end{table}

\subsection{Generalizability in New Test Environments}
Table~\ref{tab:test_env_comparison} compares the test performance of the considered algorithms under two new environments with different mean INR ranges. First, in the ``Just test'' mode, the models trained under the original environment with mean INR range \([-10,5]\) dB are directly tested in the two new environments to evaluate their generalizability. Second, in the ``Train and test'' mode, the models are trained from scratch in each new environment and then tested in the same environment, providing a clearer view of the achievable performance when the agents are adapted to the new channel and interference conditions.

In Environment~1, with mean INR range \(\mathcal{U}[0,15]\) dB, the interference condition is more challenging, and the link availability threshold is assumed to be \(A^{\mathrm{th}}=0.9\). In the ``Just test'' mode, SAC and TD3 show the best generalization among the DRL algorithms, achieving link availability values of \(0.94\) and \(0.906\), respectively, while maintaining consecutive-outage probabilities around \(10^{-2}\), which are close to the MR performance with availability \(0.953\) and consecutive-outage probability \(0.0099\). However, TD3 requires relatively high resource consumption compared to the other RL algorithms, while SAC requires less resources than DDPG and TD3. A2C consumes the least resources among the DRL methods, but its link availability is very low, indicating that its low resource usage comes at the expense of reliability. When the algorithms are trained and tested directly in Environment~1, the performance improves for several methods. In particular, PPO achieves the highest link availability among the DRL algorithms, \(0.952\), with a consecutive-outage probability of \(0.0099\), but this comes at the cost of nearly maximum resource consumption. SAC, DDPG, and TD3 also meet the availability threshold, while DDPG improves its performance compared with the direct-transfer case.

In Environment~2, with mean INR range \(\mathcal{U}[-15,0]\) dB, the interference condition is less severe, and the reliability performance improves significantly. In the ``Just test'' mode, SAC generalizes well, achieving link availability values of \(0.995\); it is the only algorithm that satisfies the availability constraint while providing a particularly favorable trade-off with \(2.6\times10^4\) resource consumption. After training and testing directly in Environment~2, SAC maintains strong performance, achieving availability of \(0.996\) and a consecutive-outage probability of \(2.2\times10^{-6}\) with comparable resource consumption. TD3 also achieves a very low consecutive-outage probability after retraining, with the highest link availability close to MR. DDPG, PPO, and A2C show larger variability and lower availability than the threshold.
Overall, the results show that algorithm performance is highly environment-dependent. SAC demonstrates the most consistent balance between resource consumption, consecutive-outage mitigation, and link availability across both direct-transfer and retrained settings, indicating better robustness to changes in the interference environment.

\begin{table*}[t]
\centering
\caption{Comparison of Test Results Under Two New Environments with Different INR Ranges.}
\label{tab:test_env_comparison}
\renewcommand{\arraystretch}{1.15}
\setlength{\tabcolsep}{4pt}
\resizebox{\textwidth}{!}{%
\begin{tabular}{|c|l|c|c|c||c|c|c|}
\hline
\multirow{2}{*}{\textbf{Mode}} 
& \multirow{2}{*}{\textbf{Algorithm}} 
& \multicolumn{3}{c||}{\textbf{Environment 1: Mean INR range \boldmath{\(\mathcal{U}[0,15]\) dB}}}
& \multicolumn{3}{c|}{\textbf{Environment 2: Mean INR range \boldmath{ \(\mathcal{U}[-15,0]\) dB}}} \\
\cline{3-8}
& 
& \textbf{Consumed resources} 
& \textbf{Consecutive outages} 
& \textbf{Link availability}
& \textbf{Consumed resources} 
& \textbf{Consecutive outages} 
& \textbf{Link availability} \\
\hline
\hline

\multirow{5}{*}{\rotatebox{90}{Just test}}
& SAC  
& $6.32 \times 10^{4} \pm 1.6 \times 10^{4}$ 
& $0.0154 \pm 0.0046$ 
& \boldmath{$0.94 \pm 0.01$}
& $2.6 \times 10^{4} \pm 6.6 \times 10^{3}$ 
& $2.97 \times 10^{-4} \pm 0.0013 $ 
& \boldmath{$0.994 \pm 0.0077$}\\

& DDPG 
& $7.51 \times 10^{4} \pm 2.5 \times 10^{4}$ 
& $0.0132 \pm 0.0032$ 
& $0.872 \pm 0.22$
& $3.28 \times 10^{4} \pm 2.4 \times 10^{4}$ 
& $3.32 \times 10^{-5} \pm 1.2 \times 10^{-4}$ 
& $0.878 \pm0.29 $ \\

& TD3  
& $7.9 \times 10^{4} \pm 2.2 \times 10^{4}$ 
& \boldmath{$0.0128 \pm 0.0068$}
& $0.906 \pm0.18 $
& $4.53 \times 10^{4} \pm 3.4 \times 10^{4}$ 
& \boldmath{$2.7 \times 10^{-5} \pm 1.2 \times 10^{-4}$} 
& $0.955 \pm 0.19$  \\

& A2C  
& \boldmath{$2.49 \times 10^{4} \pm 8.3 \times 10^{3}$}
& $0.0183 \pm 0.0045$ 
& $0.224 \pm0.076 $
& $2.16 \times 10^{4} \pm 8.2 \times 10^{3}$
& $4.3 \times 10^{-4} \pm 6.4 \times 10^{-4}$ 
& $0.23 \pm0.079 $ \\

& PPO  
& $2.64 \times 10^{4} \pm 6.2 \times 10^{3}$ 
& $0.0158 \pm0.0042 $ 
& $0.375 \pm 0.19$
& \boldmath{$ 1.93\times 10^{4} \pm 1.1 \times 10^{4}$} 
& $3.1 \times 10^{-4} \pm 0.0013$ 
& $0.445 \pm 0.27$ \\
\hline

\multirow{5}{*}{\rotatebox{90}{Train and test}}
& SAC  
& \boldmath{$5.2\times 10^{4} \pm  1.3\times 10^{4}$} 
& $0.0131 \pm 0.0023$ 
& $0.935\pm0.017$
& $3.08\times 10^{4} \pm 1.4\times 10^{4}$ 
& \boldmath{$6.05\times 10^{-6} \pm 1.7\times 10^{-5}$} 
& $0.996\pm0.0021$ \\

& DDPG 
& $8.03\times 10^{4} \pm 2.5 \times 10^{4}$ 
& $0.0101 \pm 2.7 \times 10^{-4}$ 
& $0.952\pm0.0018$
& $4.27\times 10^{4} \pm 3.1\times 10^{4}$ 
& $6.27\times 10^{-5} \pm 1.7\times 10^{-4}$ 
& $0.801\pm0.35$ \\

& TD3  
& $8.26\times 10^{4} \pm 2.1 \times 10^{4}$ 
& $0.0107\pm0.0018$ 
& $0.951\pm0.0037$
& $4.34\times 10^{4} \pm 2.3\times 10^{4}$ 
& $6.55\times 10^{-6} \pm  1.9\times 10^{-5}$ 
& \boldmath{$0.997\pm3.9\times 10^{-4}$} \\

& A2C  
& $5.52\times 10^{4} \pm 3.1 \times 10^{4}$ 
& $0.0122\pm0.002$ 
& $0.52\pm0.3$
& $2.3\times 10^{4} \pm 1.9\times 10^{4}$ 
& $3.92\times 10^{-4} \pm 2.6\times 10^{-4}$ 
& $0.233\pm0.19$ \\

& PPO  
& $9.81\times 10^{4} \pm 2.3 \times 10^{3}$ 
& \boldmath{$0.00998\pm6.8 \times 10^{-5}$} 
& \boldmath{$0.952\pm0.0013$}
& \boldmath{$1.51\times 10^{4} \pm 1.4\times 10^{4}$} 
& $5.78\times 10^{-5} \pm 1.8\times 10^{-4}$ 
& $0.462\pm0.3$ \\
\hline

\multirow{2}{*}{}
& MR   
& $1 \times 10^{5}$ 
& $0.00994$ 
& $0.953$
& $1 \times 10^{5}$ 
& $0$ 
& $0.997$ \\

& RA   
& $5.42 \times 10^{3}$ 
& $0.343$ 
& $0.202$
& $5.42 \times 10^{3}$ 
& $0.0481$ 
& $0.477$ \\
\hline
\end{tabular}%
}
\end{table*}

\subsection{Performance Under Different Energy Efficiency Weights}
Table~\ref{tab:weight_comparison} compares the test performance of the algorithms under two energy-efficiency weights, \(\omega=10\) and \(\omega=0.1\). For \(\omega=10\), the energy-efficiency term is strongly emphasized, leading most DRL algorithms to reduce resource consumption compared to \(\omega=1\) (cf. Fig.\ref{fig.res}(b)). However, this reduction is accompanied by a significant degradation in link availability for DDPG, TD3, A2C, and PPO. SAC is the only DRL algorithm that satisfies availability, achieving \(0.991\) while consuming approximately \(32\%\) of the maximum resources. This indicates that SAC is more robust in balancing reliability and resource efficiency when energy efficiency is highly prioritized.

When \(\omega\) is reduced to \(0.1\), the reliability penalties become more dominant, and most algorithms increase their resource usage to improve reliability. DDPG and TD3 significantly reduce their consecutive-outage exceedance probabilities and achieve link availability close to \(0.99\), but with higher resource consumption and large variability. A2C and PPO also achieve high availability; however, they consume nearly the maximum resource budget, making them close to the MR scheme and therefore inefficient. In contrast, SAC achieves high availability of \(0.991\) and a low consecutive-outage exceedance probability of \(8.34\times10^{-5}\), while consuming only \(37\%\) of the resources. These results demonstrate that SAC provides the most effective trade-off between reliability and resource efficiency across the considered energy-efficiency weights.

\begin{table*}[t]
\centering
\caption{Comparison of Test Results Under Different Energy Efficiency Weights.}
\label{tab:weight_comparison}
\renewcommand{\arraystretch}{1.15}
\setlength{\tabcolsep}{4pt}
\resizebox{\textwidth}{!}{%
\begin{tabular}{|l|c|c|c||c|c|c|}
\hline
\multirow{2}{*}{\textbf{Algorithm}} 
& \multicolumn{3}{c||}{\textbf{Weight of Energy Efficiency \boldmath{$\omega = 10$}}} 
& \multicolumn{3}{c|}{\textbf{Weight of Energy Efficiency \boldmath{$\omega = 0.1$}}} \\
\cline{2-7}
& \textbf{Consumed resources} 
& \textbf{Consecutive outages} 
& \textbf{Link availability}
& \textbf{Consumed resources} 
& \textbf{Consecutive outages} 
& \textbf{Link availability} \\
\hline
\hline

SAC  
& $3.28 \times 10^{4} \pm 6.9 \times 10^{3}$
& \boldmath{$7.95 \times 10^{-5} \pm 4.5 \times 10^{-5}$} 
& \boldmath{$0.991 \pm 0.0012$}
& \boldmath{$3.75 \times 10^{4} \pm 9.2 \times 10^{3}$} 
& $8.34\times 10^{-5} \pm 4.4\times 10^{-5}$
& $0.991 \pm0.0015 $ \\

DDPG 
& $2.79 \times 10^{4} \pm 2.8 \times 10^{4}$ 
& $0.0817 \pm 0.083$ 
& $0.457 \pm 0.42$
& $4.99 \times 10^{4} \pm 2.5 \times 10^{4}$ 
& $6.64 \times 10^{-5} \pm 4 \times 10^{-5}$ 
& $0.992 \pm 4.1\times 10^{-4} $ \\

TD3  
& $2.85 \times 10^{4} \pm 2.9 \times 10^{4}$ 
& $0.14 \pm 0.12 $ 
& $0.432 \pm 0.42$
& $5.74 \times 10^{4} \pm 2.4 \times 10^{4}$ 
& $5.98 \times 10^{-5} \pm 3.3 \times 10^{-5}$ 
& $0.992 \pm 8.9\times 10^{-4} $ \\

A2C  
& $1.78 \times 10^{4} \pm 1.9 \times 10^{4}$ 
& $0.137 \pm 0.032$ 
& $0.185 \pm 0.19 $
& $9.5 \times 10^{4} \pm 1 \times 10^{4}$ 
& $4.7 \times 10^{-5} \pm 2.8 \times 10^{-5}$ 
& $0.993 \pm 0.0012 $ \\

PPO  
& \boldmath{$1.77 \times 10^{4} \pm 4 \times 10^{3}$} 
& $4.83 \times 10^{-4} \pm 2.3\times 10^{-4} $ 
& $0.227 \pm0.025 $
& $9.41 \times 10^{4} \pm 2.8 \times 10^{3}$ 
& \boldmath{$4.18 \times 10^{-5} \pm 1.5 \times 10^{-6}$} 
& \boldmath{$0.993 \pm 1.7\times 10^{-4} $} \\
\hline
\end{tabular}%
}
\end{table*}

\section{Conclusion}\label{sec.con}
This paper addressed the challenge of ensuring reliability in URLLC service for coexisting in-factory subnetworks. 
IIoT applications demand intelligent and risk-sensitive approaches, especially in dynamic wireless environments affected by random channel conditions and interference, to meet strict reliability and latency requirements while maximizing resource efficiency. This motivated our research to develop a link adaptation (LA) solution leveraging reinforcement learning (RL) capable of adapting to these uncertainties to mitigate consecutive outages while minimizing resource wastage.
To this end, we proposed a novel approach to overcome the temporal structure of transmission failure using the soft actor-critic (SAC) algorithm. The proposed method dynamically adjusts transmission power and blocklength based on delayed channel quality, number of recent consecutive outages, and experienced link availability, achieving a balance between reliability and energy efficiency (EE). We evaluated the proposed SAC-based approach against other deep reinforcement learning (DRL) baselines, including DDPG, TD3, A2C, and PPO.
The results demonstrated that the SAC algorithm achieves superior performance, effectively balancing consecutive outage reduction and energy consumption, albeit with the longest inference time. 
It was also shown that SAC provides the best generalization among the DRL algorithms with a favorable trade-off between resource consumption and reliability.
Additionally, we showed that adjusting the weights of the objective function allows for fine-tuning between energy consumption and reliability performance. While DDPG and TD3 showed moderate performance and potential as alternatives to SAC, offering a balance between computational cost and performance. 
The other baselines, A2C and PPO, with the shortest inference times, exhibited poor performance, underscoring the need for advanced resource allocation and LA strategies in dynamic wireless communication environments.
This study highlights the potential of RL-based approaches, particularly SAC, in achieving optimal reliability and resource efficiency in IIoT scenarios, providing a robust framework for future advancements in Industry 4.0 and beyond.

\vspace{-6pt}
\bibliographystyle{IEEEtran}
\bibliography{reference.bib}

\end{document}